\documentclass[twocolumn,showpacs,amsmath,amssymb]{revtex4}

\usepackage{graphics}
\usepackage[dvips]{graphicx}

\begin{document}

\title[Cosmological simulations using a static scalar-tensor theory]{Cosmological 
simulations using a static scalar-tensor theory}

\author{M A Rodr\'\i guez-Meza$^{1*}$,
A X Gonz\'alez-Morales$^2$, R F Gabbasov$^1$, and Jorge L Cervantes-Cota$^1$}

\address{$^1$Depto. de F\'{\i}sica, Instituto Nacional de Investigaciones
Nucleares, Col. Escand\'on, Apdo. Postal 18-1027, 11801 M\'{e}xico D.F.}
\address{$^2$Departamento Ingenierias,
Universidad Iberoamericana,
Prol. Paseo de la Reforma 880
Lomas de Santa Fe, Mexico D.F.
Mexico \\
$^*$mar@nuclear.inin.mx; http://astro.inin.mx/mar
}

\date{\today}

\begin{abstract}
We present $\Lambda$CDM $N$-body cosmological simulations in the framework
of a static general  scalar-tensor theory of gravity.  Due to the influence of
the non-minimally coupled scalar field, the gravitational potential is modified by a Yukawa 
type term, yielding a new structure formation dynamics. We present some preliminary results and, in particular, 
we compute  the density and velocity profiles of the most massive group.
\end{abstract}

\maketitle

\section{Introduction}
The problem of explaining the structure formation of the Universe is one of the most 
fascinating at the beginning of this new millennium. From the recent experimental 
developments and observations, cosmology has acquired the status of a 
high precision area of research \cite{Breton2004}. 
In fact, recent and independent observational data measured in the CMBR
on various angular scales \cite{Bennet2003,deBernardis2000},
in type Ia supernovae \cite{Filippenko2004},
as well as in galaxy surveys \cite{Efstathiou2002,Peacock2002},
suggest that 
$\Omega=\Omega_{\Lambda}+\Omega_m \approx 1$, 
or that $\Omega_{\Lambda}\approx 0.73$ and 
$\Omega_m \approx 0.27$, 
implying the existence of dark energy and dark matter, 
respectively. However, the nature of dark components is still unknown.  
These observations favour the now standard $\Lambda$CDM 
cosmological model. 
Despite some problems on galactic scales revealed by numerical simulations, which could be explained 
with some alternative models (e.g. Warm Dark Matter),
large scale structure simulations, predicted by the $\Lambda$CDM model,
agree well  with observations.
Naturally, particular inflationary scenarios motivated by different particle 
physics theories have their own dark matter candidates, such as axion, neutralino,
Higgs particle, among others,  as well as a quintessence 
field \cite{Caldwell1998,Macorra2004}.
In order to include such particles in cosmological models 
one usually adds a scalar field (SF)
equation  to the general relativity field equations. One possibility is to couple this field non-minimally to gravity 
to have a scalar-tensor theory of gravity (STT).

In this work we present some results of the role played by a massive 
non-minimally coupled SF on the $\Lambda$CDM universe structure formation process.  Our theoretical 
model is built from a general STT static SF  which modifies the cosmological dynamics and the 
Newtonian gravity potential on particles; the dynamical full treatment of SF perturbations is now under 
development.  We evolve a cosmological cube using standard 
$\Lambda$CDM equations  with periodic  boundary conditions, where the particles 
interact through the Newtonian force plus an additional term.  The latter comes from a Yukawa 
type potential derived from the Newtonian limit of a STT.  It should be noted that in the 
present work the SF does not  replace the dark matter or dark energy, 
but rather coexists with them. To perform the simulations we have modified a standard serial 
treecode developed by one of us (MARM) \cite{Gabbasov2006} and Gadget 1 \cite{Springel2001} 
in order to take into account the contribution of the Yukawa potential.

\section{Evolution equations using a static STT}
In a previous paper we found the solutions for the potential-density pair problem in
the Newtonian limit of a scalar-tensor theory of gravity \cite{mar2004}. Here, we applied
those results and found that the potential of a single particle of mass $m$ is given by \cite{CeRoNu07}
\begin{equation} \label{phi_New}
\Phi_N = -\frac{G_N}{1+\alpha} 
\frac{m}{r}
 (1+ \alpha \, e^{-r/\lambda}) \, ,
\end{equation}
For local scales, $r \ll \lambda$, deviations from the Newtonian
theory are exponentially suppressed, and for $r \gg \lambda$ the
Newtonian constant diminishes (augments) to $G_{N}/(1+\alpha)$ for
positive (negative) $\alpha$.   This means that equation (\ref{phi_New}) 
fulfills all local tests of the Newtonian dynamics, and it is only constrained 
by experiments or tests on scales larger than --or of the order of-- $\lambda$, which in our case 
is of the order of galactic scales.  

To simulate cosmological systems,  the expansion of the Universe has to be
taken into account. Here, we employ a cosmological model with a static scalar field which is consistent with the 
Newtonian limit given by Eq. (\ref{phi_New}). Thus, the scale factor, $a(t)$,  is given by the following Friedman model,
\begin{equation} \label{new_friedman}
a^3 H^2= H_{0}^{2} \left[\frac{\Omega_{m0} +  \Omega_{\Lambda 0} \, a^3}{1+\alpha} 
+  \left(1-\frac{\Omega_{m 0}+\Omega_{\Lambda 0}}{1+\alpha} \right) \, a  \right]
\end{equation}
where $H=\dot{a}/a$,  $\Omega_{m 0}$ and $\Omega_{\Lambda 0}$ 
are the matter and energy density evaluated at present, respectively.   We notice that the source of the cosmic evolution is deviated by the term $1+\alpha$ when compared to the standard Friedman-Lemaitre 
model. Therefore, it is convenient to define a new density parameter by 
$\Omega_i^{\alpha} \equiv \Omega_i/(1+\alpha)$. This new density parameter is such that 
$\Omega_m^{\alpha} + \Omega_\Lambda^{\alpha} =1$, which implies a flat Universe, and this shall be assumed 
in our following computations, where we consider $(\Omega_m^{\alpha}, \Omega_\Lambda^{\alpha}) = (0.3, 0.7) $.  For positive values 
of $\alpha$, a flat cosmological model demands to have a factor $(1+\alpha)$ more energetic 
content ($\Omega_m$ and $ \Omega_\Lambda$) than in standard cosmology. On the other hand, for negative values of  $\alpha$ one needs a factor $(1+\alpha)$  less $\Omega_m$ and $ \Omega_\Lambda$ to have a flat Universe.  To be consistent 
with the CMB  spectrum and structure formation numerical 
experiments, cosmological constraints must be applied on $\alpha$ in order for it to 
be within the range $(-1,1)$ \cite{Nagata2002,Nagata2003,Shirata2005,Umezu2005}.

In the Newtonian limit of STT of gravity, 
the Newtonian motion equation  for a particle $i$ is written as
\begin{equation} \label{eq_motion}
\ddot{\mathbf{x}}_i + 2\, H \, \mathbf{x}_i = -\frac{1}{a^3} \frac{G_N}{1+\alpha} \sum_{j\ne i} \frac{m_j (\mathbf{x}_i-\mathbf{x}_j)}
{|\mathbf{x}_i-\mathbf{x}_j|^3} \; F_{SF}(|\mathbf{x}_i-\mathbf{x}_j|,\alpha,\lambda)
\end{equation}
where $\mathbf{x}$ is the comovil coordinate, and  the sum includes all  periodic images of particle $j$,  and $F_{SF}(r,\alpha,\lambda)$ is
\begin{equation}
F_{SF}(r,\alpha,\lambda) = 1+\alpha \, \left( 1+\frac{r}{\lambda} \right)\, e^{-r/\lambda}
\end{equation}
which,  for small distances compared to $\lambda$,  is 
$F_{SF}(r<\lambda,\alpha,\lambda) \approx 1+\alpha \, \left( 1+\frac{r}{\lambda} \right)$ and, for long 
distances, is  $F_{SF}(r>\lambda,\alpha,\lambda) \approx 1$, as in Newtonian physics. 

We now analyze the general effect that the constant $\alpha$ has on the dynamics.  The role of $\alpha$ in our approach is
as follows.   On one hand, to construct a flat model  
we have set the condition $\Omega_m^{\alpha} + \Omega_\Lambda^{\alpha} =1$, which 
implies  having $(1+\alpha)$ times the energetic content of the standard $\Lambda$CDM model. This essentially means that
we have an increment by a factor of  $(1+\alpha)$ times the amount of matter, for positive values of  $\alpha$, or a 
reduction of the same factor for negative values of  $\alpha$. Increasing or reducing this amount of matter affects 
the matter term on the  r.h.s. of the equation of 
motion (\ref{eq_motion}), but the amount affected cancels out with the term $(1+\alpha)$ in the denominator of  (\ref{eq_motion}) stemming from the new Newtonian potential (\ref{phi_New}).   On the other hand, the factor $F_{SF}$ augments (diminishes) for positive (negative)  values of $\alpha$ for small distances compared to  $\lambda$, resulting in more (less) structure formation for positive (negative) values of $\alpha$ compared to the $\Lambda$CDM model.  For $r\gg \lambda$ the dynamics is essentially Newtonian.

\section{Results}
We now present results for the $\Lambda$CDM model of the Universe model previously described. 
Because the visible component is the smaller one and given our interest to
test the consequences of including a SF contribution to the evolution equations,
our model excludes gas particles, but all its mass has been added to the dark matter. Therefore, our model 
is as follows. We start our simulation with an initial distribution of $N=2\times 32^3$ particles 
in a box with sides of $50 h^{-1}$ Mpc at $z=10$. 
This case is similar to the one that comes with Gadget 1 \cite{Springel2001}.
At present epoch, $\Omega_{b0}^{\alpha} = 0$, $\Omega_{m0}^{\alpha} = 0.3$, $\Omega_{\Lambda 0}^{\alpha}=0.7$,
$H_0= 100 h$ km/s/Mpc, $h=0.7$. We restrict the values of $\alpha$ to the interval $(-1,1)$ 
  \cite{Nagata2002,Nagata2003,Shirata2005,Umezu2005}  and  use $\lambda=5$ Mpc, since 
this scale turns out to be an intermediate scale between the size of the clump groups and the separation of 
the formed groups.

\begin{figure}
\includegraphics[width=3.1in]{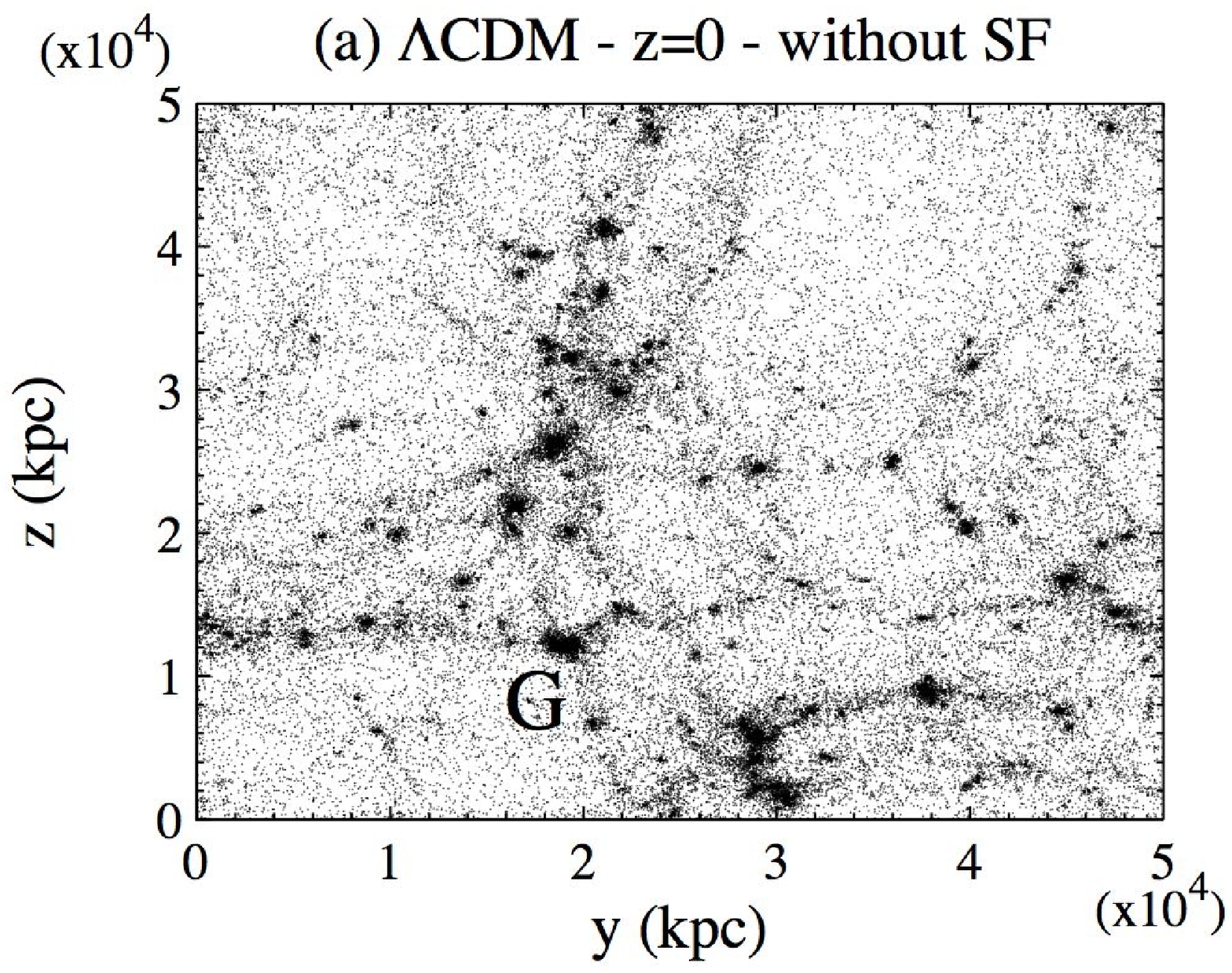}
\includegraphics[width=3.2in]{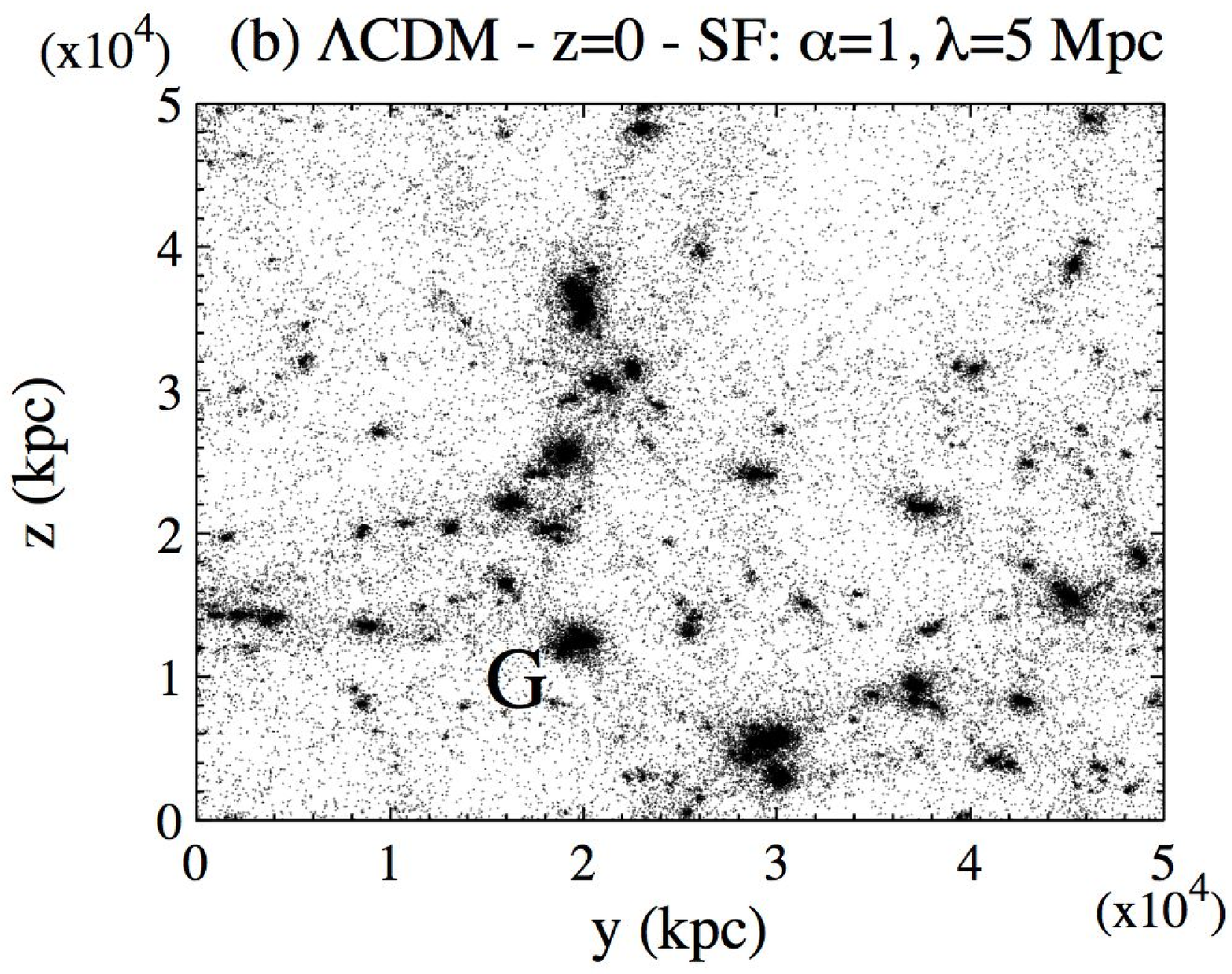}
\includegraphics[width=3.2in]{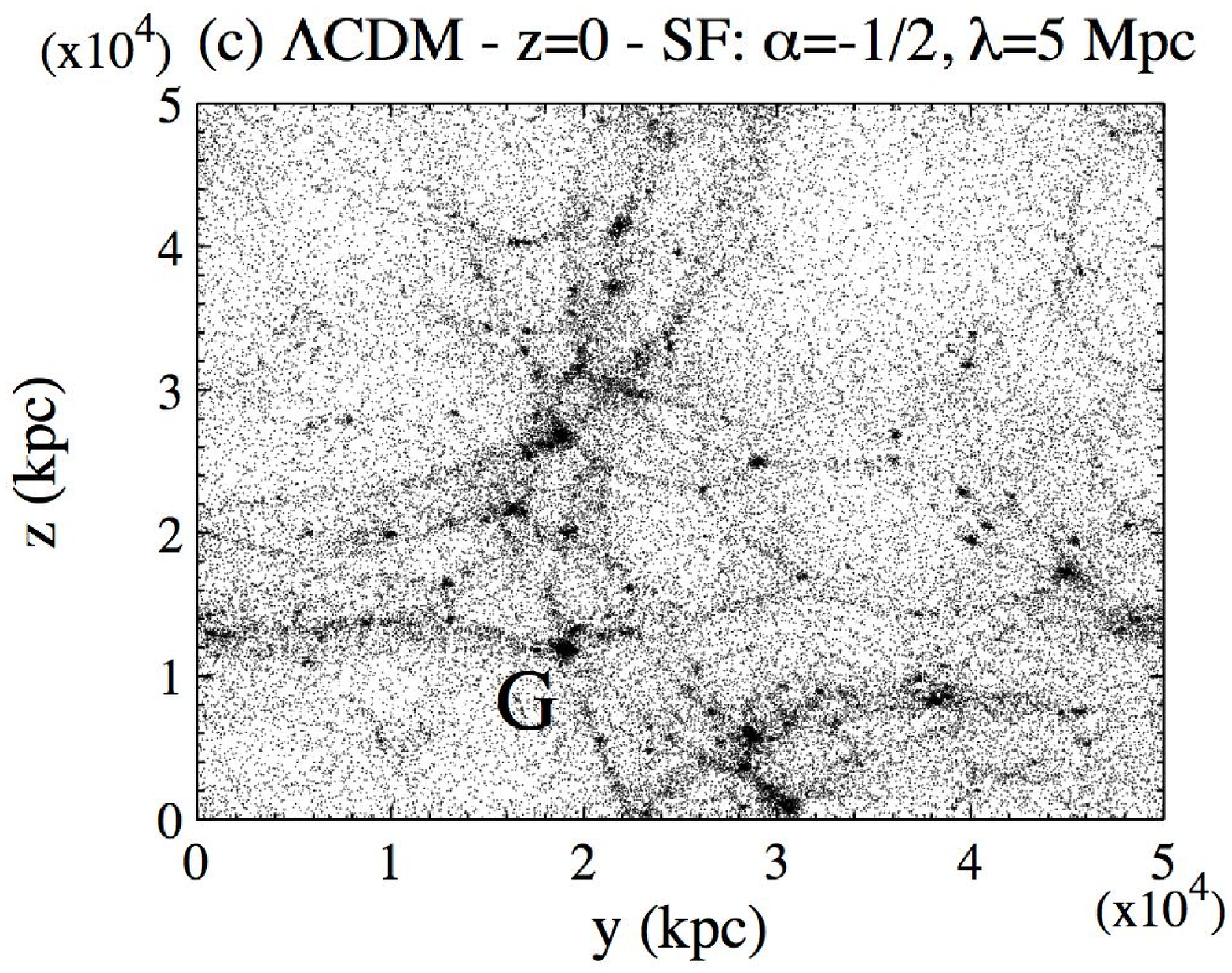}
\includegraphics[width=3.2in]{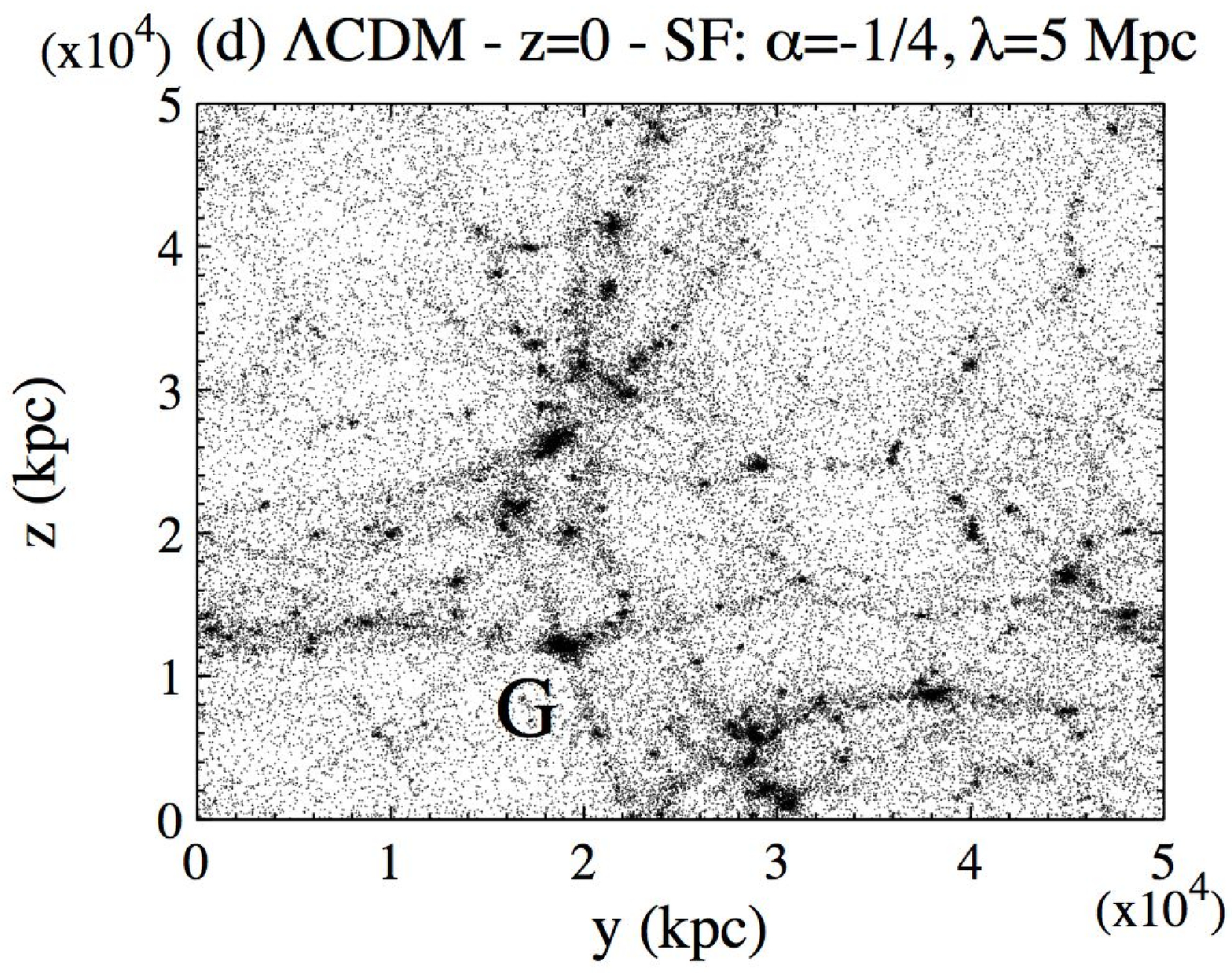}
\caption{$y$--$z$ snapshots at $z=0$ of a $\Lambda$CDM universe.  See text for details. 
}
\end{figure}
In Fig. 1 we show $y$--$z$ snapshots at redshift $z=0$ of our  $\Lambda$CDM model. 
Fig. 1 (a) presents the standard case without SF, i.e., the interaction between bodies  through
the standard Newtonian potential.
In (b) we show the case with $\alpha=1$, $\lambda=5$ Mpc.
In (c)  $\alpha=-1/2$, $\lambda=5$ Mpc.
In (d) $\alpha=-1/4$, $\lambda=5$ Mpc.  
One notes clearly how the SF modifies the matter  structure of the system. The most
dramatic cases are (b) and (c) where we have used 
$\alpha=1$ and $\alpha=-1/2$, respectively. 
Given the argument  at the end of last section, in the case of (b), for $r \ll \lambda$,
the effective gravitational pull has been  augmented by a factor of 2, 
in contrast to case (c) where it has diminished  by a factor of 1/2; in model (d) the pull 
diminishes only by a factor of 3/4. That is why one observes for $r < \lambda$ more structure 
formation in (b), less in (d), and lesser in  model (c).  The effect is  then, for a growing positive $\alpha$, to speed up  
the growth of perturbations, then of halos and then of clusters, whereas negative $\alpha$ values ($\alpha \rightarrow -1$) 
tend to slow down the growth. 

\begin{figure}
\includegraphics[width=3.1in]{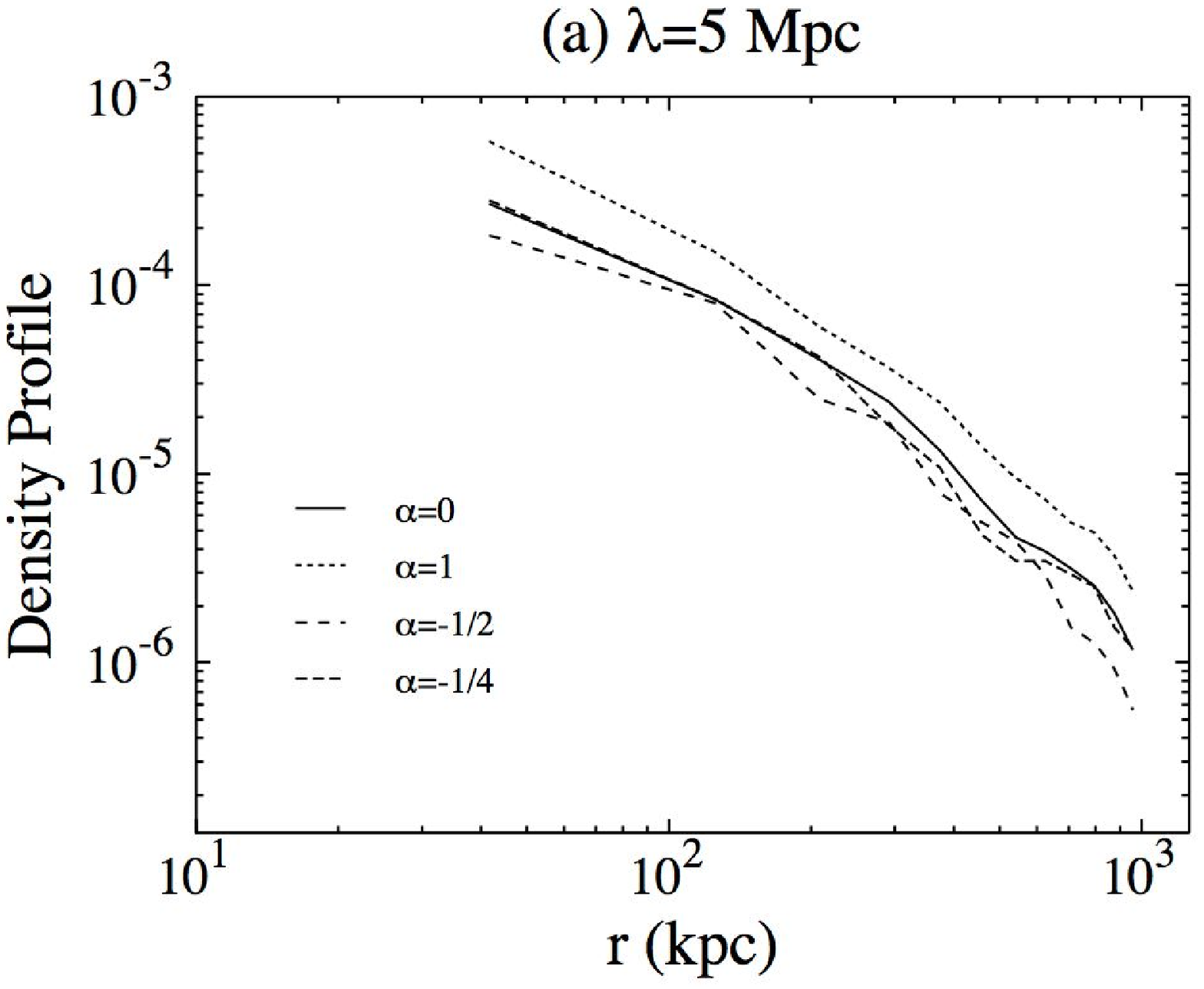}
\includegraphics[width=3.1in]{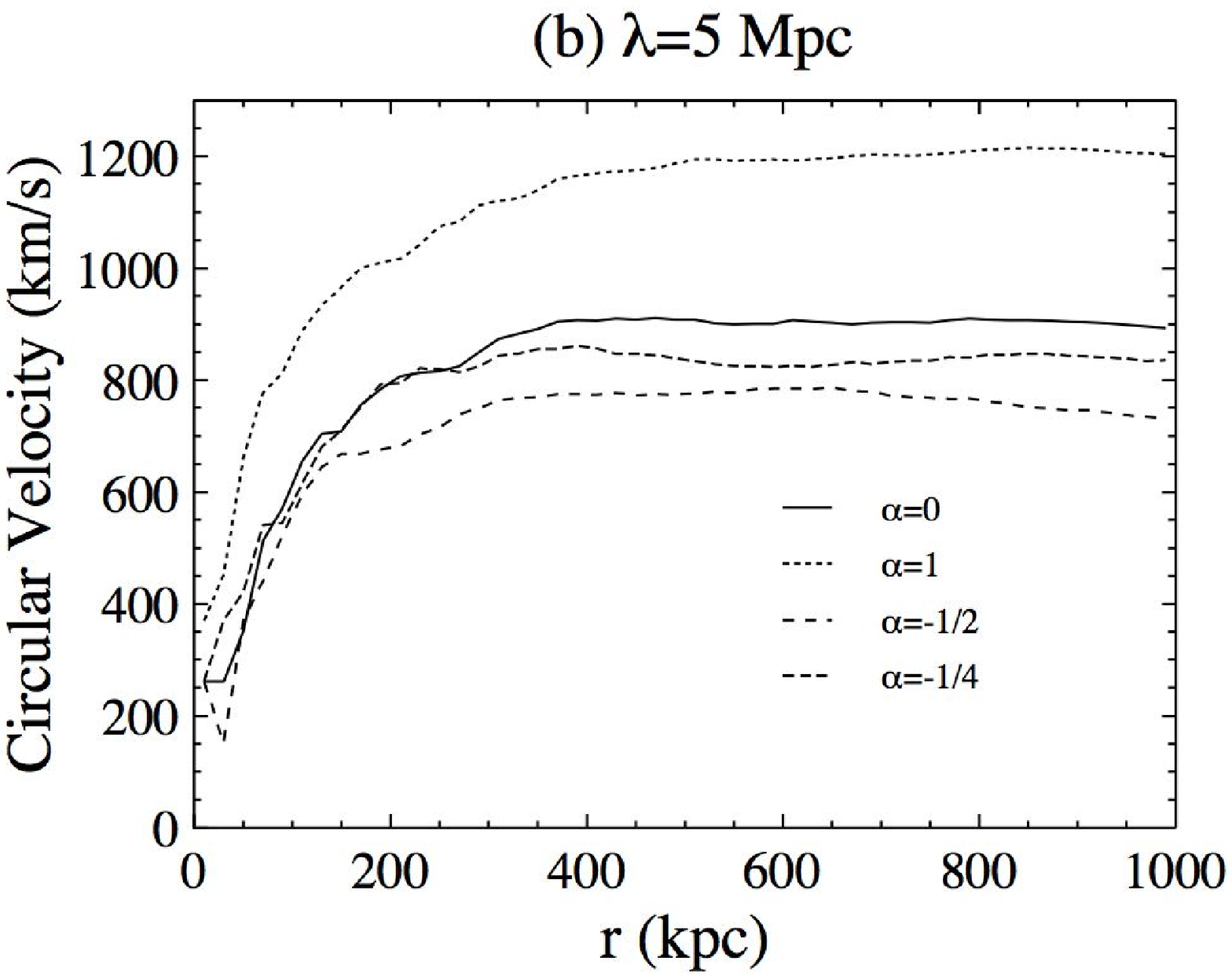}
\caption{(a) Density profiles for one of the most massive groups 
at $z=0$ of a $\Lambda$CDM universe. 
 The group is located approximately at $y=19$ Mpc, $z=12$ Mpc,
labeled with ``G'' in Fig 1(a). 
Vertical scale is in units of $\rho_0=10^{10} \mbox{M}_\odot h^{-1} / (h^{-1}\mbox{kpc})^3$.
(b) The corresponding circular velocity.
}
\end{figure}
Next, we found the groups in the system using a friend-of-friend algorithm
and select one of the most massive ones. The chosen group is located
approximately at $y=19$ Mpc, $z=12$ Mpc, and it is labeled with the letter ``G''
in Fig.\ 1(a).
The group was analyzed by obtaining their density profiles
and circular velocities.
In Fig. 2(a) we show the density profiles for this group.
The  more cuspy case is for $\alpha = 1$ and
the less cuspy is for $\alpha=-1/2$. 

In Fig. 2(b) we show, for the same group, circular velocity curves, computed using 
$v_c^2=G_N M(r)/r$. The case with $\alpha=1$ corresponds to higher values of $v_c$, since this 
depends on how much accumulated mass there is at a distance $r$ and this is enhanced by the factor 
$F_{SF}$ for positive values of $\alpha$.

The groups are at most 2 Mpc in size at $z=0$, which means that the inner structure is such that $r < \lambda$, being affected by 
the factor $F_{SF}$ as explained at the end of section 2. While for $r > \lambda$ the overall structure formation process is governed by Newtonian physics, which is the reason why the overall structure of the models in Fig. 1 is similar.

\section{Conclusions}
We have used a general, static  STT that is compatible with local observations 
by the appropriate definition of the background field constant, i.e. 
$<\phi>  =  G_{N}^{-1} (1+\alpha)$.  A direct consequence of  
our approach is that  the amount of matter (energy) has to be increased 
for positive values of $\alpha$ and diminished  for negative values of $\alpha$ 
with respect to the standard $\Lambda$CDM model 
in order to have a flat cosmological model. Quantitatively, our model demands to 
have $\Omega/ (1+\alpha) =1$ and this changes the amount of dark matter and 
energy of the model for a flat cosmological model, as assumed.   

The general gravitational effect is that  the interaction including  the SF changes by a factor 
$F_{SF}(r,\alpha,\lambda) \approx 1+\alpha \, \left( 1+\frac{r}{\lambda} \right)$ for $r<\lambda$ in 
comparison with the Newtonian case. Thus, for $\alpha >0$ the growth of structures speeds up  
in comparison with the Newtonian case.  For the   $\alpha <0$ case the effect is to diminish 
the formation of structures.  For $r> \lambda$ the dynamics is essentially Newtonian.

Using  the resulting modified dynamical equations,  we have studied the 
structure formation process of a $\Lambda$CDM universe.   We varied the amplitude and sign of the 
strength of the SF ($\alpha$) in the interval (-1,1) and performed several 3D-simulations with the 
same initial conditions.    From our simulations with different values of $\alpha$, we have found that the 
inclusion of SF changes local dynamical properties of the most massive group considered, and accordingly 
the density profile and circular velocity, however, the overall structure is somewhat similar.  Here, we notice that  
we have also studied other massive groups, in particular one smaller group located
approximately at $y=16.5$ Mpc, $z=22$ Mpc in Fig. 1(a). The trends are quite 
similar to the most massive group, so that our conclusions prevail.

In this work we only varied the amplitude of the SF ($\alpha$) leaving the 
scale length ($\lambda$) of the SF unchanged. After some preliminary runs, the increase of $\lambda$ 
enhances the structure formation process for $\alpha$ positive, and  the decrease of   $\lambda$ 
makes the structure grow at a slower rate. In a future work we will study   
an ampler space parameter and with a much better resolution. Also, the 
cosmological initial conditions will be constructed using the   
matter density field corresponding to the modified gravity.

\bigskip
{\it Acknowledgements: }
This work was supported by CONACYT, grant number I0101/131/07 C-234/07, IAC collaboration. 
The simulations were performed in the UNAM HP cluster {\it Kan-Balam}.


\end{document}